\documentclass[aps,prx,reprint,superscriptaddress,longbibliography,amsmath,amssymb]{revtex4-1}
\usepackage{slashed}
\usepackage{graphicx}
\usepackage{mathtools}
\usepackage{hyperref}
\usepackage{physics} 
\usepackage{float}
\usepackage{braket}
\usepackage{xcolor}
\usepackage{newtxtext,newtxmath}
\usepackage[T1]{fontenc}

\begin{document}


\date{\today}

\newcommand{\pcc}{\,cm$^{-3}$}	
\newcommand{\singletS}{${}^1$S$_0\,$}
\newcommand{\singletP}{${}^1$P$_1\,$}
\newcommand{\tripletS}{${}^3$S$_1\,$}
\newcommand{\tripletPj}{$^3$P$_J\,$}
\newcommand{\tripletDj}{$^3$D$_J\,$}

\newcommand{\tripletPone}{$^3$P$_1\,$}
\newcommand{\tripletPzero}{$^3$P$_0\,$}
\newcommand{\tripletPtwo}{$^3$P$_2\,$}
\newcommand{\sPj}{$^{2s+1}$P$_j\,$}
\newcommand{\tripletDone}{$^3$D$_1\,$}
\newcommand{\tripletDtwo}{$^3$D$_2\,$}
\newcommand{\sr}{$^{87}\text{Sr}$}
\newcommand{\g}{$\ket{g}$}
\newcommand{\e}{$\ket{e}$}
\newcommand{\erecR}{\ensuremath{\mathrm{E_R^{813}}}}	
\newcommand{\erecB}{\ensuremath{\mathrm{E_R^{497}}}}	
\newcommand{\erec}[1]{\ensuremath{\mathrm{E_R^{#1}}}}	
\newcommand{\qclk}{\ensuremath{q_{\mathrm{clk}}}}	
\newcommand{\kclk}{\ensuremath{k_{\mathrm{clk}}}}	
\newcommand{\wclk}{\ensuremath{\omega_{\mathrm{clk}}}}	
\newcommand{\lclk}{\ensuremath{\lambda_{\mathrm{clk}}}}	
\newcommand{\lamp}{\ensuremath{\lambda_{\mathrm{primary}}}}	
\newcommand{\lams}{\ensuremath{\lambda_{\mathrm{secondary}}}}	
\newcommand{\uK}{\ensuremath{\upmu\mathrm{K}}}	
\newcommand{\um}{\ensuremath{\upmu\mathrm{m}}}	
\newcommand{\uHz}{\ensuremath{\upmu\mathrm{Hz}}}	
\newcommand{\ketS}[1]{\ensuremath{\ket{{}^1S_0,\,#1}}}	
\newcommand{\ketP}[1]{\ensuremath{\ket{{}^3P_0,\,#1}}}	
\newcommand{\braP}[1]{\ensuremath{\bra{{}^3P_0,\,#1}}}	
\newcommand{\Szero}[1]{\ket{{}^1S_0, m_F=#1}}
\newcommand{\Pzero}[1]{\ket{{}^3P_0, m_F=#1}}

\def\scititle{An Aubry--Andr\'{e} localized optical lattice clock}

\title{\bfseries \boldmath \scititle}

\author{Shuo Ma}%
\thanks{These authors contributed equally to this work.}
\author{Anke St{\"o}ltzel}%
\thanks{These authors contributed equally to this work.}
\author{Bennett Christensen}
\thanks{Present address: Department of Physics, Colorado State University, Fort Collins, CO, USA.}
\author{Zhuoli Ding}%
\author{Dhruva Ganapathy}%
\author{Andrew Householder}%
\author{Genyue Liu}%
\author{Shimon Kolkowitz}
\email{To whom correspondence should be addressed; \\
Email: kolkowitz@berkeley.edu}

\affiliation{%
 Department of Physics, University of California, Berkeley, CA 94720, USA
}%



\begin{abstract}
Optical lattice clocks in space could improve satellite navigation systems and enable new tests of fundamental physics, including gravitational-wave detection and tests of relativity.
In microgravity, atoms can tunnel to degenerate neighboring lattice sites, broadening the clock transition and limiting coherence times.
We demonstrate Aubry–Andr\'{e} localization in a bichromatic optical lattice formed by two magic-wavelength lasers, suppressing tunneling in a horizontal lattice clock emulating free fall.
We recover a Fourier-limited clock Rabi spectrum and measure Ramsey coherence times of up to 2\,s in shallow lattices, with up to 100-fold improvement over the delocalized regime.
These results establish Aubry--Andr\'{e} localization as a practical route to controlling tunneling in future space-based clocks, and open a path to operating terrestrial clocks at shallower lattice depths.
\end{abstract}
\maketitle


Recent advances in optical lattice clocks have enabled fractional measurement precision at the $10^{-20}$ level~\cite{bothwell_resolving_2022,Xin_multiplexedOLC} and absolute accuracies at or below the $10^{-18}$ level~\cite{mcgrew_atomic_2018,bothwell_jila_2019,aeppli_clock_2024}. Optical lattice clocks have been used to probe fundamental physics through measurements of gravitational redshifts over laboratory and geodetic scales~\cite{grotti_transportable_2018,takano_geopotential_2016,takamoto_test_2020,zheng_redshift_2023} and searches for ultralight dark matter~\cite{wcislo_darkmatter_2016,derevianko_hunting_2014,kennedy_dark_2020}.
Operating these clocks in space would extend such measurements to longer baselines and to gravitational potentials unavailable on Earth~\cite{derevianko_space_clock_2022, bongs_soc_2015, origlia_space_2018,xia_css_2025}.
Networks of space-based clocks could also detect and localize gravitational waves in a frequency band between those accessible to ground-based and proposed space-based laser interferometers~\cite{kolkowitz_gravitational_2016,dhruva_GWdetection_2026_arxiv}.

A central challenge of operating optical lattice clocks in space is preventing the atoms from tunneling between lattice sites. In terrestrial clocks with a vertical lattice, gravity tilts the lattice potential, providing energy offsets between neighboring sites and thereby suppressing tunneling through Wannier--Stark localization~\cite{gluck_wannier_stark_2002, lemonde_wolf_tunneling_2005, aeppli_wannier_stark_2022, ferrari_bloch_2006}.
In microgravity, as experienced in free fall in orbit or in deep space, the lattice sites are degenerate, and atoms can freely tunnel between them at low lattice depths~\cite{lemonde_wolf_tunneling_2005}.
When the clock laser wavelength is not an integer multiple of the lattice spacing, this tunneling broadens the clock spectrum, reducing the excitation contrast and limiting the coherence of the clock~\cite{Shimon_SOC,livi_synthetic_2016,wall_soc_2016}.

Several approaches can mitigate atomic tunneling.
The simplest is to increase the lattice depth, which suppresses tunneling but requires more laser power, placing restrictions on the achievable size, weight, and power of the clock, while shallower lattices are also preferable because they reduce higher-order lattice light shifts~\cite{brown_hyperpolarizability_2017} and Raman scattering~\cite{jack2025}.
Rotating the satellite can provide an effective acceleration \cite{space_odyssey,clement2007artificial} that would result in Wannier-Stark localization, but this approach is incompatible with applications such as gravitational wave detection and measurements of relativity that require the clock to be in an inertial reference frame~\cite{armano_lpf_2018}.
An accordion lattice~\cite{li_accordion_2008,williams_accordion_2008} could make the transition frequency independent of quasimomentum by choosing the beam angle such that the lattice spacing is an integer multiple of the clock wavelength, while also suppressing tunneling by increasing the distance between neighboring sites.
To our knowledge, this approach has not yet been demonstrated in an optical lattice clock, and it reduces the number of trapping sites and the trap depth per unit laser power while requiring precise control of the beam geometry.
Periodically shaking the lattice can induce dynamical localization~\cite{dunlap_kenkre_1986,lignier_dynamical_2007,shaken_OLC_exp_paper_2025,shaken_OLC_exp_paper_2022,shaken_OLC_blochSiegert_shift2026}, but introduces additional systematic shifts, micromotion-induced spectral distortions, and potentially larger probe shifts.\\
In this work, we propose and experimentally demonstrate an alternative method for suppressing tunneling in a shallow optical lattice, without requiring moving parts or time-dependent modulation.
We realize Aubry--Andr\'{e} (AA) localization~\cite{aubry_andre_1980} (also sometimes referred to as a form of Anderson localization in prior literature \cite{roati_anderson_2008}) in a bichromatic optical lattice formed by two lattices at different magic wavelengths.
We implement this in a $^{87}$Sr optical lattice clock with a horizontal lattice, in which neighboring lattice sites are degenerate in energy, as they would be in microgravity.
The introduction of a weaker, secondary lattice lifts this degeneracy of the sites of the primary lattice by imposing quasiperiodic energy offsets between them, which localize the atoms.
We observe the predicted AA phase transition and a resulting recovery of a narrow, Fourier-limited clock carrier and find that Ramsey coherence times can be improved by up to two orders of magnitude.
This approach enables coherent clock spectroscopy at low lattice depths without relying on a gravitational potential, providing a route to controlling tunneling in future space-based optical lattice clocks, and potentially also allowing Earth-based clocks to operate at shallower lattice depths, reducing lattice light shifts and Raman scattering rates.

\begin{figure*}[ht!]
	\centering
	\includegraphics[scale=0.85]{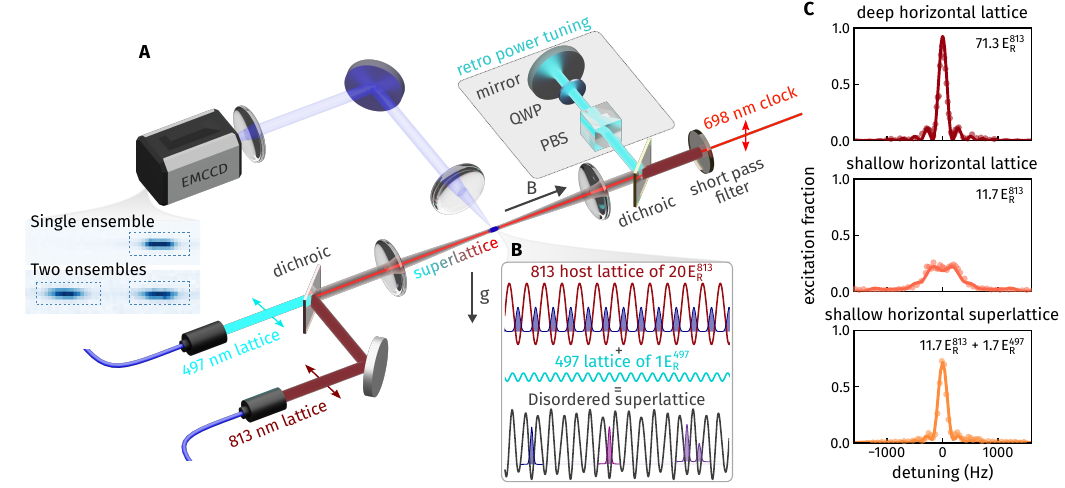} 
	\caption{\textbf{A horizontal strontium optical lattice clock with two collinear lattices at incommensurate magic wavelengths and variable depths.} \textbf{(A)} Schematic of the setup confining \sr\,in a horizontal one-dimensional optical lattice formed by 813\,nm light co-aligned with a $497\,\mathrm{nm}$ lattice with tunable power in the retro-reflected beam allowing for a dipole trap for radial confinement, and a variable secondary lattice depth. Clock laser light propagates axially along the lattice, interrogating either a single ensemble of atoms via Rabi spectroscopy, or two ensembles via differential Ramsey spectroscopy. \textbf{(B)} (Top) Potential for an 813\,nm lattice with a depth of 20\,\erecR\ in dark red, with the probability density of the delocalized ground state \sr\ atomic wavefunction shown in blue. Adding 1\,\erecB\ of 497\,nm lattice (Middle) introduces disorder to the potential (Bottom), plotted with three representative atomic probability densities, illustrating two localized wavefunctions on sites with non-degenerate neighboring sites, and one wavefunction occupying a near-degenerate pair of sites. \textbf{(C)} Experimental data, along with a fit, showing: (Top) Rabi spectrum of a single ensemble in a deep horizontal lattice at 813\,nm where tunneling is suppressed by the potential barrier height. (Middle) Rabi spectrum for atoms in a shallow horizontal lattice where tunneling is significant, resulting in broadening and suppression of excitation. (Bottom) Rabi spectrum in a shallow, disordered bichromatic superlattice exhibiting AA localization.}
	\label{fig:1} 
\end{figure*}
\section*{Experimental setup}
We experimentally realize the tunneling and localized regimes using two collinear, horizontal optical lattices near 813\,nm and 497\,nm (Fig.~\ref{fig:1}A), both tuned to their operational magic wavelengths for the \sr\ \singletS\,--\,\tripletPzero\ clock transition.
The new magic wavelength near 497\,nm was recently measured in Ref.~\cite{barreiro_497_magic_wavelength}.
We load ensembles of \sr\ atoms into a single lattice (either 813\,nm or 497\,nm), which we designate the primary lattice, and ramp up the secondary lattice only during clock spectroscopy.
At low primary lattice depths, gravity (perpendicular to the horizontal lattice axis) can cause atoms to fall out of the lattice radially.
We therefore attenuate the retroreflected 497\,nm beam for both choices of primary lattice, so that the incoming 497\,nm beam acts as a dipole trap that provides radial confinement, while the weaker retroreflection sets the depth of the 497\,nm lattice.
The atoms are cooled and prepared in the stretched state \singletS\ ($m_F=-9/2$), and interrogated with the 698\,nm clock light that propagates along the lattice axis.
We measure clock spectra using single-ensemble Rabi spectroscopy, and coherence using differential Ramsey spectroscopy~\cite{Xin_multiplexedOLC} between two ensembles separated by 1\,mm.
Throughout this work, each lattice depth $U_\lambda$ is expressed in units of its recoil energy, $\mathrm{E_R}^{\lambda}=h^{2}/(2m\lambda^{2})$, where $m$ is the atomic mass. The ratio of these units is \erecB/ \erecR\, = 2.68.
Lattice depths are calibrated using clock sideband spectroscopy.

\section*{The Aubry--Andr\'{e} Hamiltonian}
The tunneling-induced spectral broadening of the optical clock transition, and its suppression by localization, can be modeled and understood using the Aubry--Andr\'{e} (AA) Hamiltonian~\cite{aubry_andre_1980}.
In the tight-binding limit of a deep primary lattice, it is obtained by perturbatively projecting the weaker secondary lattice modulation onto the lowest band of the primary lattice~\cite{modugno_exponential_2009, roati_anderson_2008,Domínguez_AA_explanation_2019} (Fig.~\ref{fig:1}B):
\begin{align}
H_{\mathrm{AA}} = &-J \sum_i \left( \ket{w_{i+1}}\bra{w_i} + \ket{w_i}\bra{w_{i+1}} \right) \nonumber\\
&+ \Delta \sum_i \cos(2\pi\beta i+\phi) \ket{w_i}\bra{w_i}
\label{eq: AA_Hamiltonian}
\end{align}
where $\ket{w_i}$ denotes the ground-band Wannier state of the primary lattice at site $i$, $J$ is the nearest-neighbor tunneling matrix element of the primary lattice, $\Delta$ is the amplitude of the on-site energy modulation, $\beta=\lambda_{\mathrm{primary}}/\lambda_{\mathrm{secondary}}$ is the ratio of the two lattice wavelengths, and $\phi$ is the relative phase of the two lattices.\par

In the absence of the secondary lattice ($\Delta=0$), the motional eigenstates are delocalized Bloch states labeled by the dimensionless quasimomentum $q$, with dispersion $E(q)=-2J\cos q$. Driving the clock transition between the electronic states \singletS\ ($\ket{g}$) and \tripletPzero\ ($\ket{e}$) shifts the quasimomentum by $q_{\mathrm{clk}}=\pi\lambda_{\mathrm{primary}}/\lambda_{\mathrm{clk}}$, coupling $\ket{g,q}$ to $\ket{e,q+q_{\mathrm{clk}}}$~\cite{lemonde_wolf_tunneling_2005,Shimon_SOC,wall_soc_2016,livi_synthetic_2016}.
Because the ratio of clock wavelength and lattice spacing is not an integer, the clock transition frequency becomes quasimomentum-dependent.
The transition frequencies therefore span a band of width $(8J/h)\lvert\sin(q_{\mathrm{clk}}/2)\rvert$.
For an ensemble that uniformly populates the ground band (i.e. the temperature of the atoms is high enough such that all energies within the ground band are filled with approximately equal probability, but is small compared to the band gap), this produces a broadened spectrum with split peaks at the band edges associated with van Hove singularities~\cite{lemonde_wolf_tunneling_2005,Shimon_SOC,wall_soc_2016}.

Adding the secondary lattice introduces a finite $\Delta$ that competes with tunneling.
For irrational $\beta$ (i.e. incommensurate\footnote{We use incommensurate not in the strict mathematical sense of an irrational ratio, but in the practical sense that the ratio cannot be written as a ratio of integers small enough to produce a repeating pattern within the relevant system size.} wavelengths $\lambda_\mathrm{primary},\,\lambda_\mathrm{secondary}$), the ideal AA model undergoes a phase transition at $\Delta/J=2$~\cite{Domínguez_AA_explanation_2019,aubry_andre_1980}, with extended states below the threshold, and exponentially localized states above it~\cite{roati_anderson_2008,lahini_quasiperiodic_2009}.
In the localized regime, the lattice absorbs the photon recoil, so clock excitation preserves the atoms' motional state~\cite {lemonde_wolf_tunneling_2005}.
Because both lattices are at magic wavelengths for the clock transition, the site energies are identical for $\ket{g}$ and $\ket{e}$.
Together, these two conditions give all atoms the same transition frequency and restore a narrow spectrum. 

\begin{figure*}[ht!]
	\centering
	\includegraphics[scale=0.85]{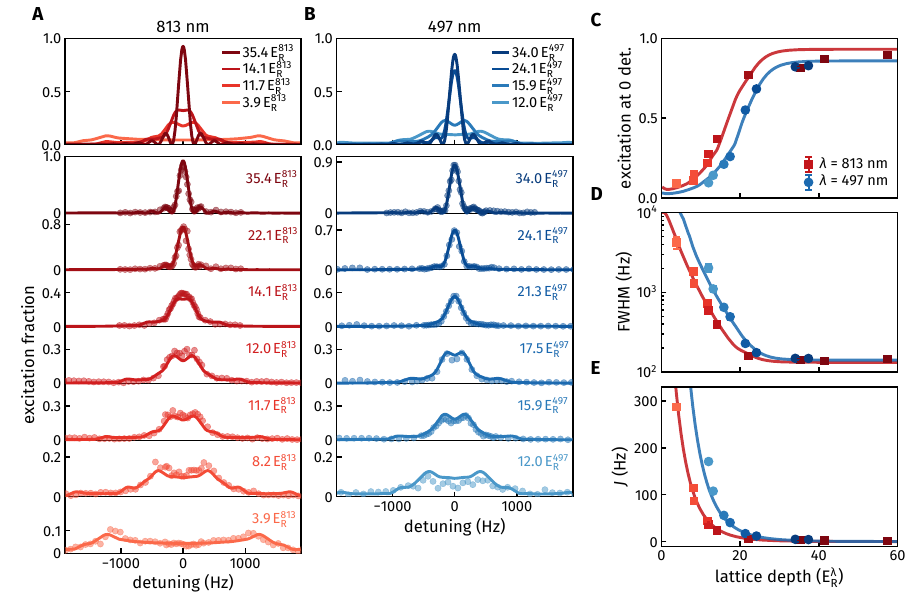}
    \caption{\textbf{Tunneling-induced broadening of Rabi spectra in individual 1D horizontal lattices, at 813\,nm and 497\,nm.}
    \textbf{(A)}\&\textbf{(B)} Excitation fraction as a function of clock laser detuning for the 813\,nm (left, red points) and 497\,nm (right, blue points) lattices at various lattice depths, following a $\pi$--pulse with $\Omega=2\pi\times104$\,Hz and $2\pi\times111$\,Hz, respectively. The theoretical curves (solid lines in matching colors) are overlaid with experimental data (round markers). The topmost panels summarize all spectra shown below on a common excitation fraction scale. 
    \textbf{(C)} On-resonance excited-state fraction versus lattice depth for the 813\,nm (red squares) and 497\,nm (blue circles) lattices.  Points are experimental data extracted from the measured spectra and solid lines show the corresponding model predictions.
    \textbf{(D)} Full width at half maximum (FWHM) versus lattice depth, extracted from Lorentzian fits to the measured and calculated spectra.
    \textbf{(E)} Tunneling rate $J$ versus lattice depth, comparing values extracted from fits to the measured spectra (points) with theoretical predictions (solid lines). Error bars on fitted quantities in \textbf{(C-E)} represent $\pm1\sigma$ uncertainties of the fitted parameters.}
    \label{fig:2}
\end{figure*}

\section*{Rabi spectra in tunneling and localized regimes}
\label{sec:Rabi}
To study the effects of tunneling in a single horizontal lattice, corresponding to the $\Delta=0$ limit of Eq.~\eqref{eq: AA_Hamiltonian}, we measure Rabi spectra of the clock transition as the lattice depth is reduced, measuring independently in an 813\,nm lattice and a 497\,nm lattice (Fig.~\ref{fig:2}A\&B).
At large depths, the spectrum shows a Fourier-limited resonance with an excitation fraction of 0.9 limited by the $\pi$--pulse fidelity.
As the lattice becomes shallower, the spectrum broadens and develops two peaks at the edge of the ground band, while the on-resonance excitation fraction decreases.
We quantify the broadening through the on-resonance excitation fraction (Fig.~\ref{fig:2}C) and an effective full width at half maximum (FWHM), extracted from Lorentzian fits to the spectra (Fig.~\ref{fig:2}D).
To extract the tunneling rate, we fit each spectrum with the lineshape of Eqs.~(S1)--(S2), averaged over the thermal radial distribution of the atoms~\cite{blatt_rabi_2009}, with $J$ as the only free parameter
(see Supplementary Materials).
The fitted tunneling rates as a function of measured lattice depth agree with the values calculated from the band structure of each lattice (Fig.~\ref{fig:2}E).

\begin{figure*}[ht!]
	\centering
	\includegraphics[scale=0.9]{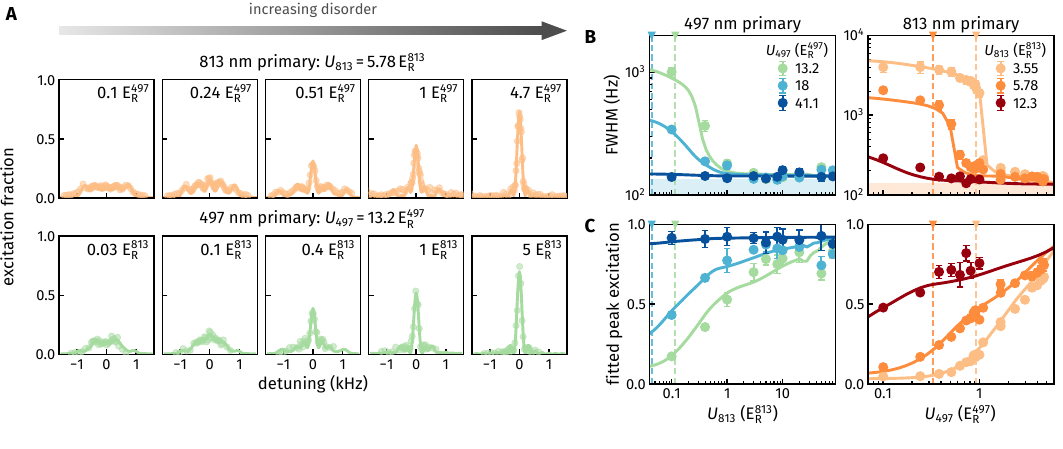} 
    \caption{\textbf{Localization induced recovery of the clock spectrum.} \textbf{(A)} Excitation fraction in \tripletPzero versus clock detuning for the 813\,nm primary lattice at 5.78\,\erecR\ (orange points, top row) and the 497\,nm primary lattice at 13.2\,\erecB\ (green points, bottom row), for increasing secondary lattice depth. The lineshape predicted by the model is shown by the solid lines in matching color. Increasing the secondary lattice depth narrows the broadened spectrum and restores the carrier. \textbf{(B)\&(C)} FWHM and on-resonance excitation, respectively, versus secondary lattice depth for different primary lattice depths. The measured Rabi spectra (points) and calculated spectra (solid lines) are fitted with a Lorentzian function to extract the effective FWHM and on-resonance excitation fraction. At higher secondary lattice depths, the FWHM saturates at the Fourier limit determined by the Rabi pulse duration (indicated by the shaded region). The vertical, dashed lines along with the colored arrows indicate the localized phase transition thresholds $\Delta/J=2$. For the deepest primary lattice depth, this threshold is below the secondary depths plotted here. Error bars represent $\pm1\sigma$ uncertainties of the fitted parameters.} 
    \label{fig:3} 
\end{figure*}

Adding the secondary lattice recovers a narrow clock transition in a shallow primary lattice, for both choices of primary lattice (Fig.~\ref{fig:3}A). As in Fig.~\ref{fig:2}, we quantify the recovery through an effective FWHM extracted from Lorentzian fits to the spectra (Fig.~\ref{fig:3}B) and the on-resonance excitation fraction (Fig.~\ref{fig:3}C).
For the shallowest primary lattice, the FWHM decreases by more than an order of magnitude, down to the Fourier limit set by the finite $\pi$--pulse duration, and the on-resonance excitation increases from below 0.1 to above 0.8.
This recovery requires only a weak secondary lattice: for the 497\,nm primary lattice at 13.2\,\erecB, the localization transition occurs at an 813\,nm lattice depth of about 0.1\,\erecR, below 1\% of the primary lattice depth.

Dashed vertical lines in Fig.~\ref{fig:3}B\&C denote the secondary lattice depth at which the AA model predicts the phase transition, $\Delta/J = 2$, for each primary lattice depth.
The clock transition narrows and the fraction of atoms excited increases as the secondary lattice depth approaches these values, consistent with the onset of AA localization.
To describe the full range of measured depths, including combinations beyond the weak-modulation regime of the AA model, we calculate the clock spectra by solving the Hamiltonian of the continuum bichromatic potential (see Supplementary Materials), which reduces to Eq.~\eqref{eq: AA_Hamiltonian} in the limit of a deep primary lattice and a weak secondary lattice~\cite{modugno_exponential_2009}.
With the atom temperatures and the $\pi$--pulse fidelity as free parameters, the calculation reproduces the recovery of the clock transition and the dependence of the linewidth and on-resonance excitation on the secondary lattice depth (Fig.~\ref{fig:3}, solid lines).
\section*{Coherence time improvement in differential Ramsey spectroscopy}
The spectral resolution of Rabi spectroscopy is limited by the clock laser linewidth.
In order to further characterize the improvement in coherence from AA localization, we employ differential Ramsey spectroscopy, in which two spatially separated atomic ensembles are interrogated simultaneously by a shared clock laser beam, so that laser phase noise is common-mode and can be rejected.
At each dark time, we scan the phase of the second Ramsey $\pi/2$--pulse and extract the contrast from an ellipse fitted to the measured excitation fractions of the two ensembles using maximum-likelihood estimation~\cite{Xin_multiplexedOLC}.
We define the coherence time as the $1/e$ decay time of an empirical bi-exponential fit to the contrast (see Supplementary Materials).
Figure~\ref{fig:4}A\&C show the contrast as a function of dark time for different combinations of primary and secondary lattice depths, and representative ellipses at a fixed dark time show the increase in contrast with secondary lattice depth in panel i\&ii.

\begin{figure*}[ht!]
	\centering
	\includegraphics[scale=0.85]{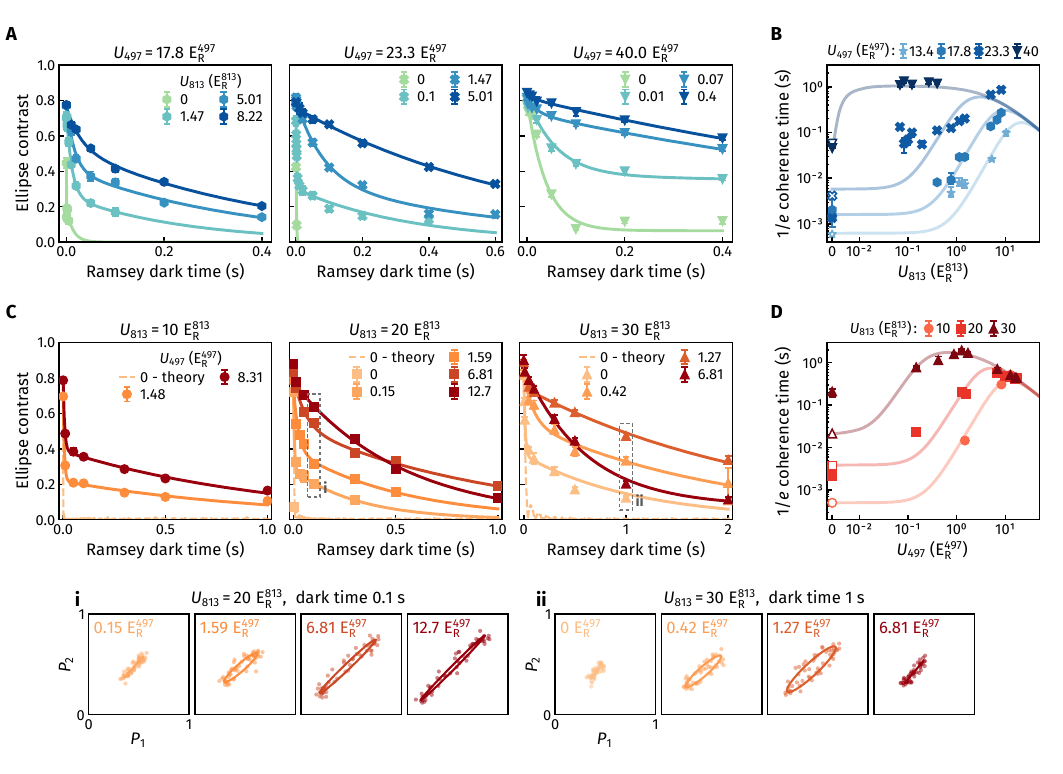}
	\caption{\textbf{
    Localization induced increase in coherence times in differential Ramsey spectroscopy.} 
    A differential Ramsey measurement repeated 20 times with varying phase between the two $\pi/2$--pulses maps out an ellipse per fixed Ramsey dark time when plotting excitation fraction in ensemble 1 parametrically over excitation fraction in ensemble 2. \textbf{(A\&C)} Measured ellipse contrast for three exemplary primary lattice depths and different disorder strength over Ramsey dark time. The contrast at a given primary lattice depth increases with disorder up to some threshold, until other decoherence effects dominate. Error bars represent the $\pm1\sigma$ uncertainties on the ellipse fit extracted from jackknifing. Solid lines are empirical bi-exponential fits to the data. \textbf{(i\&ii)} Exemplary data sets showing the improvement of ellipse contrast with disorder at a fixed Ramsey dark time of 0.1\,s for primary depth of 20\,\erecR\ and 1\,s for primary depth of 30\,\erecR. \textbf{(B\&D)} Measured $1/e$ coherence time improvement for different primary lattice depths over disorder/secondary lattice depth. Error bars represent the $\pm1\sigma$ uncertainties. Solid lines are fits to a heuristic multi-decay model to guide the eye.}
	\label{fig:4} 
    
\end{figure*}

Without the secondary lattice, the coherence is limited by tunneling, which at shallow primary lattice depths broadens the clock spectrum beyond the excitation bandwidth of the $\pi/2$--pulse (see spectra in Fig.~\ref{fig:2}) and even prevents efficient excitation of a coherent superposition across the ensemble.
In the deepest primary lattice (i.e. 30\,\erecR), we find that the measured coherence time exceeds the predictions of the calculation.
We attribute this to small residual site-to-site energy offsets, to which the dynamics are sensitive because the tunneling rate at this depth is only 1.6\,Hz.
For example, a lattice tilt of $0.1^{\circ}$ from horizontal produces an energy offset between neighboring sites comparable to the tunneling rate at this depth.
Further, the curvature of the Gaussian lattice potential one Rayleigh range away from the center produces site-to-site offsets of 8\,Hz for a 40\,\um\ beam waist at 40\,\erecR, and reflections of the 497\,nm dipole trap beam (see Experimental setup) from optical surfaces in the setup could result in a weak secondary 497\,nm lattice, and a depth of $\sim10^{-3}$\,\erecB\ would be sufficient to reach the AA localization threshold.
Because these offsets are not independently known, the dashed lines in Fig.~\ref{fig:4}C show the predicted coherence time from the model for a truly horizontal single lattice.

In Fig.~\ref{fig:4}B\&D we show the measured Ramsey coherence times in AA superlattices for primary lattice depths between 13 and 40\,\erecB and between 10 and 30\,\erecR.
The solid lines are fits to a heuristic multi-decay model to guide the eye.
For both choices of primary lattice, we observe that the measured atom-atom coherence times improve by over two orders of magnitude in the localized regime, reaching $\sim2$\,s, at which point we believe tunneling is no longer the dominant decoherence mechanism in our apparatus.
For shallower primary lattices, we observe an initial fast decay of the coherence with a characteristic timescale of $\tau_{\mathrm{fast}}\sim10$\,ms, which is not predicted by the AA model.
We speculate that this may be the result of relative phase noise between two lattices, as they are retroreflected on independent dichroic mirrors and their relative phase is not actively stabilized, but leave investigation of this phenomenon for future work. 
We find that the coherence times begin to drop again at the highest secondary lattice depths, especially when the secondary lattice depth approaches or exceeds that of the primary lattice. 
These cases are well beyond the AA regime, in which the secondary lattice only weakly shifts the on-site energies.
At comparable depths, the secondary lattice instead strongly reshapes the potential of each primary lattice site: depending on its local phase, a site's minimum can be displaced axially, its trap frequency substantially changed, or its confinement removed altogether, turning the former minimum into a local maximum.
Atoms loaded into the primary lattice are therefore projected onto a strongly site-dependent potential when the secondary lattice is ramped on, leading to motional excitation and redistribution among sites.
Consistent with this, we observe an increase in temperature in this regime, as measured by sideband thermometry.

The longest measured atomic coherence time is limited to $\sim$2\,s by technical sources of decoherence specific to our apparatus rather than by the localization method itself.
Limited magnetic field control along the transverse axes required us to orient the quantization axis parallel to the lattice axis, a configuration that maximizes vector light shifts from the lattice and dipole trap beams, and residual magnetic field gradients further dephase the two ensembles.
In addition, the available lattice power limited the loading depth, and the linewidth of our in lattice cooling light was insufficient to effectively cool to the motional ground state.
We anticipate that longer coherence times can be achieved if these problems are addressed.

We also find that, for a given primary lattice depth, the coherence recovers at a higher secondary lattice depth than predicted by the AA threshold $\Delta/J=2$ for atoms in the ground band. We attribute this to the finite temperature of the atoms ($\sim1$\,$\uK$\ radially and axially): atoms in higher axial bands tunnel much faster and therefore require a larger $\Delta$ to localize, and atoms at larger radial positions experience a shallower primary lattice and thus a larger $J$.

\section*{Discussion}
Having demonstrated that AA localization provides narrow clock transitions and long coherence times in shallow horizontal lattices, we now compare the costs and benefits of this approach with the simpler, traditional alternative of suppressing tunneling in a deep, single lattice. 

One cost of the AA technique is that the secondary lattice introduces new
potential sources of systematic uncertainty relative to a standard optical
lattice clock.
First, atoms on different sites experience different total lattice
intensities and therefore site-dependent hyperpolarizability shifts.
Because the secondary lattice is shallow and the hyperpolarizability shift
scales quadratically with lattice depth, we estimate this contribution to be less than $10^{-3}$ of that from the primary lattice at the optimal operating condition discussed below, up to the as yet undetermined ratio of the hyperpolarizability coefficient for the 497\,nm lattice.
Second, for a thermal occupation of the motional states, off-resonant transitions to other motional states may contribute unequally on opposite sides of the carrier, which can produce line pulling.
We estimate this effect in theory and find the shift on the order of several millihertz for the coherent Rabi spectrum. (See Supplementary Materials for details.)

The secondary lattice also contributes to Raman scattering. 
We measure the Raman scattering rate of the 497\,nm lattice to be $99(6)\times 10^{-4}\,\mathrm{(E_R^{497}\,s)^{-1}}$, which, to our knowledge, has not been reported previously. 
For the 813\,nm lattice, we measured Raman scattering rate of $46(2)\times 10^{-5}\,\mathrm{(E_R^{813}\,s)^{-1}}$ in a prior work \cite{jack2025}.

To compare the two approaches, we calculate the maximum coherence time set by the balance between tunneling and Raman scattering, using the calculated tunneling rates, the Raman scattering rates above, and the \tripletPzero lifetime, and assuming atoms cooled to the motional ground band (Fig.~\ref{fig:optimal_coherence} and Supplementary Materials).
In a single 813\,nm lattice in microgravity, the maximum coherence time is 36.6\,s at a depth of 100\,\erecR \ (Fig.~\ref{fig:optimal_coherence}A).
For a lattice waist of 100\,\um, this operating point requires 1.2\,W of 813\,nm laser power, assuming lossless retroreflection.
With the addition of the 497\,nm secondary lattice, the maximum coherence time increases to 116.6\,s at lattice depths of 16\,\erecR and 0.13\,\erecB \ (Fig.~\ref{fig:optimal_coherence}B), and the required powers are reduced to 190\,mW at 813\,nm and 0.8\,mW at 497\,nm, substantially reducing the optical power required for long coherence times.
%
\section*{Summary and Outlook}
In this work, we have demonstrated the suppression of tunneling in an optical lattice clock through Aubry--Andr\'{e} localization in a bichromatic lattice formed by two lattices at different magic wavelengths.
In a horizontal lattice, where the lattice sites are degenerate in energy, a weak secondary lattice restores a narrow, Fourier-limited clock transition and increases the differential Ramsey coherence time by up to two orders of magnitude, with a maximum of approximately 2\,s.
This coherence time is currently limited by technical sources of dephasing unrelated to the localization, such as residual magnetic field gradients and vector light shifts, while the fast initial decay observed in shallow primary lattices may arise from relative phase noise between the two lattices, subject to future study.
Given our results and the projected coherence times, we anticipate that Aubry--Andr\'{e} localization provides a viable path to controlling tunneling in future space-based optical lattice clocks while reducing lattice laser power requirements, albeit at the cost of requiring an additional laser system.
Aubry--Andr\'{e} localization could also improve the performance of terrestrial vertical optical lattice clocks by enabling operation at shallower lattice depths while maintaining localized atomic wavefunctions and suppressing s-wave scattering between atoms on neighboring sites~\cite{aeppli_wannier_stark_2022}.
At the same time, these results show that optical lattice clocks provide a useful new platform for studying the physics of Aubry--Andr\'{e} localization itself.

\section*{Acknowledgments}
We thank Jeff Thompson, David Weld, Jun Ye, Ed Marti, Ross Hutson, and Sara Campbell for fruitful and inspiring discussions. We thank Andrew Christensen and Anna Dardia for their feedback on the manuscript. We thank Atom Computing for their generous donation of laser systems that were critical to completing this work. We thank Jonathan Dolde for his help and advice during the assembly of the experimental apparatus. 

\paragraph*{Funding:}
This work was supported by a Packard Fellowship for Science and Engineering, a Sloan Fellowship, the Miller Institute for Basic Research in Science, the ``Table-top experiments for fundamental physics'' program sponsored by the Alfred P. Sloan Foundation, Gordon and Betty Moore Foundation, John Templeton Foundation, and Simons Foundation, the Gordon and Betty Moore Foundation under grant DOI 10.37807/GBMF12966, NASA under grant No.~80NSSC24K1561, and the National Science Foundation under Grants No.~2143870, 2326810, and the NSF QLCI Challenge Institute for Quantum Computation, award No.~2553595.
\paragraph*{Usage of generative AI:}
During the preparation of this manuscript, the authors used ChatGPT-5.4 OpenAI and Claude Opus 5 Anthropic. The authors used these models to assist with language refinement and editing for clarity. The authors generated all original concepts and text, reviewed and edited all tool-generated outputs, and take full responsibility for the accuracy, integrity, and originality of the final content. The authors also used the models mentioned above to assist in optimizing simulation speed and drafting data analysis code. All AI-assisted code was manually reviewed, debugged, and validated by the authors. No synthetic data or unverified automated processing was used in the research findings.
\paragraph*{Author contributions:}
SM and AS built the experimental apparatus together with BC, performed the measurements, analyzed the data, prepared the figures, and wrote the manuscript.
SM performed the numerical simulations.
ZD contributed to the measurements and data analysis, developed the theoretical model of line pulling and wrote the corresponding section of the manuscript.
DG, AH, and GL provided technical support and contributed to discussions.
The whole project was conceived of and overseen by SK.
All authors contributed to editing the manuscript.
\paragraph*{Competing interests:}
There are no competing interests to declare.
\paragraph*{Data, code and materials availability:}
The data that support the findings of this article are openly available at Zenodo \cite{data}.

\clearpage

\bibliography{AL_clock_paper_v2}
\bibliographystyle{apsrev4-2-custom}

\newpage

\renewcommand{\thefigure}{S\arabic{figure}}
\renewcommand{\thetable}{S\arabic{table}}
\renewcommand{\theequation}{S\arabic{equation}}
\renewcommand{\thepage}{S\arabic{page}}
\setcounter{figure}{0}
\setcounter{table}{0}
\setcounter{equation}{0}
\setcounter{page}{1}
\begin{center}
\section*{Supplementary Materials for\\\scititle}
\end{center}
\subsection*{Experimental methods}
The \sr\ atomic beam from the oven (AOSense Rev.D) is first cooled and captured in a magneto-optical trap (MOT) operating on the \singletS$\rightarrow$\singletP transition at 461\,nm. The cooling light is generated by a home-built injection-locked diode laser (Nichia NDB4916), seeded by a Toptica DL Pro stabilized using saturated-absorption spectroscopy.

The atoms are then transferred to a second-stage broadband red MOT operating on the \singletS$\rightarrow$\tripletPone$(F=11/2)$ transition at 689\,nm.
The 689\,nm light is generated by an injection-locked amplifier (MOGLabs ILA), seeded by an external cavity diode laser (MOGLabs cateye ECDL).
In the final stage of cooling, the broadband modulation is removed and the atoms are cooled to approximately 1\,\uK\ using single-frequency light near resonance.
We prepare the ensembles for the clock spectroscopy by first spin-polarizing the atoms to $\ket{^1S_0, m_F = -9/2}$ by driving the transition \singletS$\rightarrow$\tripletPone($F=9/2$) and then perform Doppler cooling in the radial direction and resolved sideband cooling in the axial direction on \singletS$\rightarrow$\tripletPone($F=11/2$).

For the measurements presented in the main text, the atoms are cooled and loaded into either the 813\,nm lattice, generated by a CW fiber laser (Precilaser), or the 497\,nm lattice, generated by a VALO SF vertical external-cavity surface-emitting laser (Vexlum), depending on the choice of primary lattice.
The two lattices are linearly polarized, combined on dichroic mirrors, and independently retroreflected from separate dichroic mirrors mounted on a monolithic optical breadboard.
The lattice beams are coaligned and have similar waists of approximately 40\,\um\ at the center of the chamber.
A constant magnetic field of 6.3\,G applied along the lattice axis defines the quantization axis.
The lattice depths are calibrated using clock-transition sideband spectroscopy at high power and extrapolated linearly in power to lower depths. Both lattice wavelengths are stabilized to a HighFinesse WS/7-30 wavemeter. 

Several methods are used to control the lattice depths.
For measurements using the 497\,nm lattice as the primary lattice, its depth is adjusted by directly varying the incident lattice power, with an additional OD~0.15 neutral-density filter attenuating the retroreflected beam.
The depth of the secondary 813\,nm lattice is similarly controlled by varying its incident power.
For low 813\,nm lattice depths ($<$3\,\erecR), the 813\,nm laser power is stabilized at a higher operating point and an additional neutral-density filter is inserted to passively reduce the power delivered to the atoms.

For measurements using the 813\,nm lattice as the primary lattice, the incident 497\,nm power is held fixed at 350\,mW as a (running-wave) dipole trap.
The depth of the secondary 497\,nm standing wave is then varied by adjusting the retroreflected power using a polarizing beam splitter and a quarter-wave plate on a motorized rotation stage, as shown in Fig.~\ref{fig:1}.
This allows the modulation depth to be changed while keeping the incident 497\,nm power fixed.
The depths of both lattices are calibrated using clock sideband spectroscopy. For the 813\,nm primary lattice, we measure the sidebands over depths from 21\,\erecR\ to 71\,\erecR\ and extrapolate the calibration to lower lattice powers.
We detect the \singletS population using fluorescence imaging on the resonant 461\,nm \singletS$\rightarrow$\singletP transition, with the fluorescence collected on an EMCCD camera (Andor iXon Ultra 888).
Population remaining in the \tripletPzero state is measured by first repumping the atoms to the ground state using 679\,nm and 707\,nm (AOSense) lasers, followed by a second fluorescence image. Finally we take a final image as a background measurement. 

The 698\,nm clock light is generated by a home-built injection-locked diode laser seeded by a Menlo clock laser system. The clock light is linearly polarized and propagates coaxially with the optical lattices.
Its frequency is controlled by tuning the RF frequency applied to an acousto-optic modulator.
The experimental sequence is controlled by a Quantum Machines OPX+.

The clock laser drifts by approximately 1--2\,Hz/s over the course of a day. To compensate for this drift, we interleave a frequency-tracking sequence with the measurements presented in the main text, and use the measured clock resonance to recenter the probe frequency.

\subsection*{Operational magic wavelength measurement}
All measurements are performed with both optical lattices tuned to their respective operational magic wavelengths for the clock transition used in the measurement. For measurements with the 497\,nm lattice as the primary lattice, we interrogate the $\Szero{-9/2}\leftrightarrow\Pzero{-7/2}$ transition.

For measurements with the 813\,nm lattice as the primary lattice, we use the $\Szero{-9/2}\leftrightarrow\Pzero{-7/2}$ transition for Rabi spectroscopy and the $\Szero{-5/2}\leftrightarrow\Pzero{-3/2}$ transition for differential Ramsey spectroscopy.
The latter transition has reduced magnetic sensitivity and therefore improves the atom-atom coherence.
\begin{figure}[ht!]
    \centering
    \includegraphics[width=0.95\linewidth]{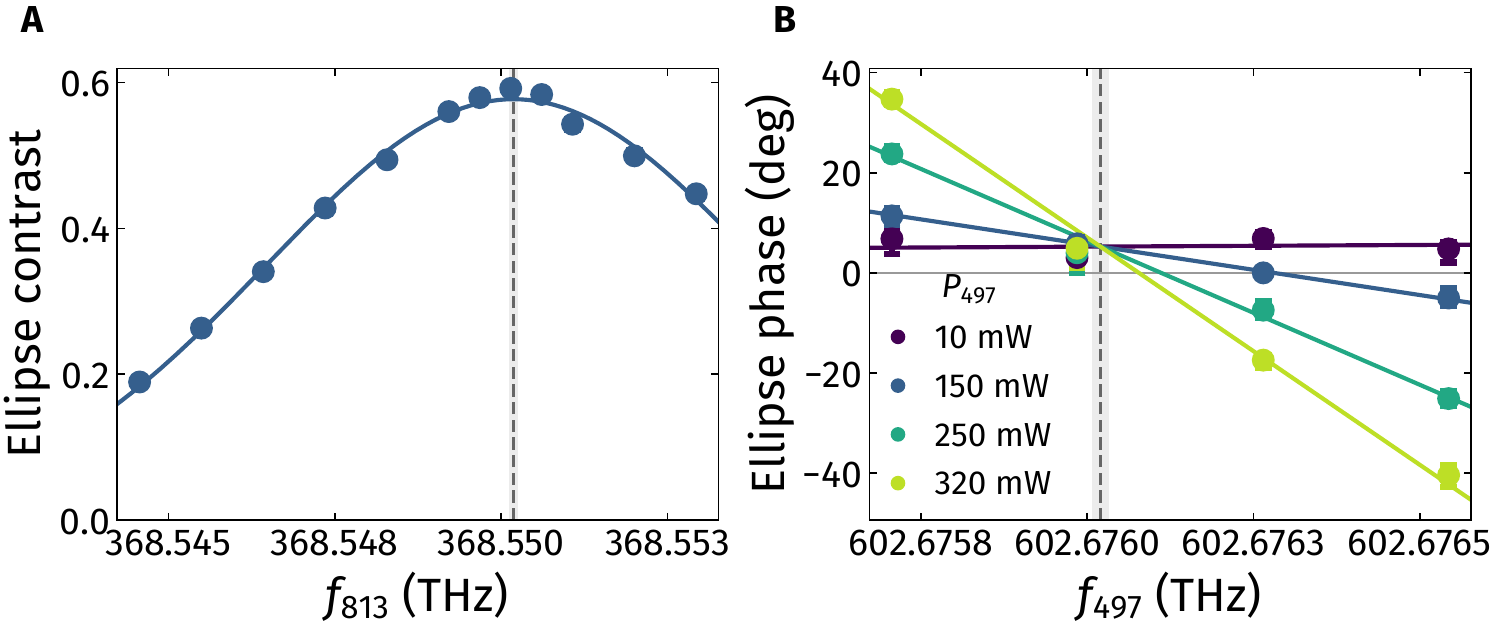}
    \caption{\textbf{Measurement of magic wavelengths near 497\,nm and 813\,nm for the $\ket{{}^1S_0, m_F=-5/2}\leftrightarrow\ket{{}^3P_0, m_F=-3/2}$ clock transition.} \textbf{(A)} Ramsey contrasts in the 813\,nm lattice over lattice frequency at fixed Ramsey dark time of 0.3\,s. The data is fitted to a Gaussian (solid line). Its maximum indicates the magic frequency (dashed line). \textbf{(B)} Ellipse phase over 497\,nm lattice frequency at fixed Ramsey dark time of 0.5\,s for four different powers of the 497\,nm lattice. The measurement is performed in the bichromatic lattice with the 813\,nm lattice deep and at its magic wavelength. The dashed line shows the measured magic frequency of 497\,nm, where the ellipse phase is independent of 497\,nm light intensity. }
    \label{fig:S1}
\end{figure}
For the $\ket{{}^1S_0, m_F=-5/2}\leftrightarrow\ket{{}^3P_0, m_F=-3/2}$ transition, we determine the operational magic wavelength of the 813\,nm lattice following the method presented in Ref.~\cite{Xin_multiplexedOLC}.
We prepare two spatially separated ensembles in an 813\,nm lattice with a depth of 72\,\erecR\ and perform synchronous Ramsey spectroscopy with a fixed dark time of 0.3\,s.
For each lattice wavelength, we fit the excitation fractions of the two ensembles to an ellipse using maximum likelihood estimation and extract the Ramsey contrast.
The contrast is maximized when the differential light shift vanishes.
From this measurement, we obtain an 813\,nm magic frequency of 368.550439(69)\,THz (Fig.~\ref{fig:S1}(A)).

For the 497\,nm lattice, we use a related approach.
In a single 497\,nm lattice, the large tunneling rate limits the Ramsey coherence time (Fig.~\ref{fig:4}).
We therefore perform the measurement in the bichromatic lattice, with the 813\,nm lattice operated at its magic wavelength, where the atoms are localized and the Ramsey coherence time is extended.
We fix the Ramsey dark time to 0.5\,s. Because the two ensembles experience slightly different 497\,nm intensities, any residual differential light shift appears as a differential clock frequency between them.
We scan the 497\,nm lattice frequency over a range of 1\,GHz and measure the differential clock frequency at several 497\,nm lattice depths.
At the magic frequency, the differential shift becomes independent of the 497\,nm lattice depth. From this measurement, we obtain a 497\,nm magic frequency of 602.676066(16)\,THz (Fig.~\ref{fig:S1}(B)).

For the $\ket{{}^1S_0, m_F=-9/2}\leftrightarrow\ket{{}^3P_0, m_F=-7/2}$ transition, the magic wavelengths are shifted relative to those of the  $\ket{{}^1S_0, m_F=-5/2}\leftrightarrow\ket{{}^3P_0, m_F=-3/2}$  transition by the $m_F$-dependent tensor and vector polarizabilities of the \tripletPzero state, where the vector contribution arises from residual ellipticity of the lattice polarization.
Using similar differential Ramsey measurements, we obtain magic frequencies of 368.54539(38)\,THz for the 813\,nm lattice and 602.675204(29)\,THz for the 497\,nm lattice.\par

Comparing our measured operational values with the previously reported magic frequency from Ref.~\cite{barreiro_497_magic_wavelength} for the 497\,nm lattice on the  $\ket{{}^1S_0, m_F=\pm9/2}\leftrightarrow\ket{{}^3P_0, m_F=\pm9/2}$  $\pi$-transitions, 602.6750(5)\,THz, and for the 813\,nm lattice on the $\ket{{}^1S_0, m_F=+5/2}\leftrightarrow\ket{{}^3P_0, m_F=+3/2}$ transition from Ref.~\cite{Xin_multiplexedOLC}, 368.554810(30)\,THz, our measured values for both lattices are shifted well beyond the standard errors, which we attribute primarily to the vector light shift. Previous precision measurements of the magic wavelength have typically used a quantization field perpendicular to the lattice propagation direction, a geometry that strongly suppresses the vector light shift. In the measurements performed here, the quantization field was aligned along the lattice axis due to the geometric constraints of our apparatus. This geometry modifies the $m_F$-dependent tensor contribution and maximizes the sensitivity to the vector light shift, such that residual ellipticity of the lattice polarization can produce a substantially larger $m_F$ dependent shift of the operational magic wavelength.

\subsection*{Raman scattering rate in the 497\,nm lattice}
We measure the Raman scattering rate of the 497\,nm lattice by comparing the lifetimes of atoms prepared in either the \singletS or \tripletPzero state, following the methods employed in Refs.~\cite{dorscher2018} and~\cite{jack2025}.
Raman scattering of lattice photons transfers atoms out of \tripletPzero, which appears as an additional loss relative to \singletS.
We load the atoms into the 497\,nm lattice, prepare them in either \singletS or \tripletPzero, and measure the atom number remaining in that state after a variable hold time, for several lattice depths.
We calibrate each lattice depth with sideband spectroscopy and convert it to the effective trap depth.
The average atom number is approximately 400 total atoms held over an estimated 2500 lattice sites, so we neglect density-dependent on-site collisional loss.
\begin{figure}[ht!]
    \centering
    \includegraphics[width=0.55\linewidth]{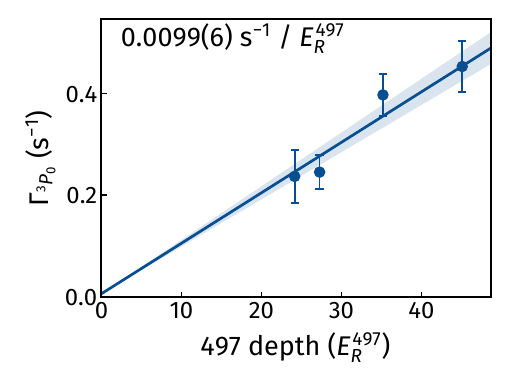}
    \caption{\textbf{Raman scattering in the 497\,nm lattice.} Additional loss rate $\Gamma$ of \tripletPzero atoms relative to \singletS atoms as a function of the 497\,nm lattice depth. The line is a linear fit with the offset fixed at the \tripletPzero decay rate $1/\tau$, $\tau=167$\,s. The shaded band shows the $1\sigma$ uncertainty of the fitted slope.}
    \label{fig:raman}
\end{figure}
We fit each decay with an exponential, $N(t)=A\,e^{-\gamma t}$, separately for \singletS ($\gamma_g$) and \tripletPzero ($\gamma_e$).
We assume that loss mechanisms common to both states, such as background gas collisions and heating, contribute equally to $\gamma_g$ and $\gamma_e$, and take the difference $\Gamma=\gamma_e-\gamma_g$ as the additional loss of \tripletPzero. We attribute $\Gamma$ to Raman scattering and to the spontaneous decay of \tripletPzero, and fit $\Gamma = R\,U_{497} + 1/\tau$ with the natural lifetime fixed at $\tau=167$\,s~\cite{jack2025}.
We obtain a Raman scattering rate of $R=0.0099(6)\,\mathrm{s^{-1}}$ per \erecB\ (Fig.~\ref{fig:raman}).
At the secondary lattice depth of $0.13\,\erecB$ used in Fig.~\ref{fig:optimal_coherence}, this corresponds to $1.3\times10^{-3}\,\mathrm{s^{-1}}$.

\subsection*{Theoretical modeling}
\subsubsection*{Simulation for Fig.~\ref{fig:2}}
We simulate the clock Rabi spectra in the absence of the secondary lattice following the approach previously used to study the dynamics of spin-orbit coupled atoms in the optical lattice clock~\cite{Shimon_SOC}.
We assume that the atoms populate the lowest Bloch band of the primary lattice uniformly in quasimomentum.
For each $q$, the clock laser couples the initial Bloch state
$\ket{{}^1S_0, n_z = 0, q}$ to the excited Bloch state $\ket{{}^3P_0, n_z = 0, q + q_{\mathrm{clk}}}$ with the Rabi frequency $\Omega$
and the quasimomentum-dependent detuning
\begin{equation}
  \delta_q = \delta
  - \left[ E_e(q + q_{\mathrm{clk}}) - E_g(q) \right] / \hbar,
\end{equation}
where $\delta$ is the laser detuning from the bare clock transition and
$E_{g}$ and $E_{e}$ are the band energies in the ground and excited states.
The excitation probability for a pulse of duration $t$ is
\begin{equation}
  P_q(\delta) = \frac{\Omega^2}{\Omega^2 + \delta_q^2}
  \sin^2\!\left( \frac{\sqrt{\Omega^2 + \delta_q^2}\, t}{2} \right),
\end{equation}
and the spectrum is obtained by averaging $P_q(\delta)$ over $q$ in the
first Brillouin zone.

To account for the finite radial temperature, we further average the
spectra over the thermal radial distribution of atoms.
Atoms at different
radial positions sample different local lattice depths because of the
Gaussian profile of the lattice beams.
The simulations shown in
Fig.~\ref{fig:2} assume radial temperatures between $1$ and
2\,\uK.

\subsubsection*{Simulation for Fig.~\ref{fig:3}}
To model the clock spectra over the range of quasiperiodic modulation
explored experimentally, we numerically solve the full axial Hamiltonian
\begin{equation}
H = \frac{\hat{p}^{2}}{2m}
- U_{497}\sin^{2}(k_{497}z + \phi)
- U_{813}\sin^{2}(k_{813}z),
\label{eq:full_hamiltonian}
\end{equation}
where $m$ is the atomic mass, $U_{497}$ and $U_{813}$ are the lattice depths,
and $\phi$ is the relative phase between the two lattices. Because both
lattices are magic, the ground and excited clock states experience the same
potential.

To make the calculation tractable while retaining the spatial structure of
the quasiperiodic potential, we approximate the wavelength ratio by a nearby rational number,
\begin{equation}
\beta = \frac{\lambda_{813}}{\lambda_{497}} \simeq \frac{18}{11},
\end{equation}
such that $N_{497} = 18$ periods of the 497\,nm lattice approximately
coincide with $N_{813} = 11$ periods of the 813\,nm lattice.
This approximation replaces the incommensurate bichromatic potential by a periodic supercell.
The Hamiltonian then recovers translational symmetry over the supercell, and its eigenstates can be described in a reduced Brillouin zone.

We expand the axial wavefunction in plane-wave states $\ket{j}$, with $j = -M, \ldots, M$.
For a quasimomentum $q \in [-1, 1)$ in the reduced Brillouin zone, in units of $k_{813}/N_{813}$, the momentum of each basis state is
\begin{equation}
p_j = \hbar\frac{(q + 2j)\,k_{813}}{N_{813}}
    = \hbar\frac{(q + 2j)\,k_{497}}{N_{497}},
\label{eq:basis}
\end{equation}
In units of $E_R^{813}$, the Hamiltonian matrix is
\begin{equation}
H_{jj'} = \left(\frac{q + 2j}{N_{813}}\right)^{2}\delta_{jj'}
- \frac{\beta^{2} s_{497}}{4}\,\delta_{j,j'\pm N_{497}}
- \frac{s_{813}}{4}\,\delta_{j,j'\pm N_{813}},
\label{eq:hamiltonian}
\end{equation}
where $s_i = U_i/E_R^{i}$ is the depth of each lattice in its own recoil energy, and $E_R^{497}/E_R^{813} = 2.68 = \beta^{2}$ and $\phi$ is set to 0.

For each quasimomentum, we diagonalize the Hamiltonian to obtain the eigenstates and eigenenergies.
Initial motional states are sampled according to a Boltzmann distribution.
For the internal state, we assume atoms are in the ground state \singletS.
For each initial state, we calculate the coupling to all final motional states.

The final spectrum is obtained by averaging over the thermally populated axial states and over the radial distribution of atoms, which accounts for the variation of the lattice depths across the finite transverse extent of the atomic ensemble. In the simulation, we use a radial temperature of 1\,\uK.
The axial temperature in the superlattice is not independently known and is therefore treated as a fit parameter. For each fixed primary lattice depth in Fig.~\ref{fig:3}B\&C, we use a single axial temperature shared across the corresponding curve as the secondary lattice depth is varied.
The fitted axial temperatures range from 0.3 -- 0.5\,\uK\ for the 813\,nm primary lattice and are approximately 1.5\,\uK\ for the 497 nm primary lattice.
Although the axial temperature in the superlattice is not independently measured, the fitted temperatures follow the same trend with primary lattice depth as those extracted from the spectra measured in the corresponding host lattice alone at the same depth.
To account for the finite $\pi$--pulse fidelity, we scale all calculated spectra by a single factor for each primary lattice, determined from the spectrum measured in the deepest single lattice, where tunneling is negligible.

\subsection*{Data analysis methods}
The clock Rabi spectroscopy lineshapes associated with delocalized atomic wavefunctions below or near the AA phase transition threshold are complex, with multiple peaks. To quantify the spectral narrowing across the localization transition, we fit each measured Rabi spectrum with the Lorentzian function
\begin{equation}
  P_e(\delta) = \frac{A\,(\Gamma/2)^2}{(\delta-\delta_0)^2 + (\Gamma/2)^2},
  \label{eq:lorentzian}
\end{equation}
where $A$ is the peak excitation, $\delta_0$ is the line center, and $\Gamma$ is the FWHM, and use the extracted FWHM $\Gamma$ as the metric for comparison.
To obtain the on-resonance excitation fraction, we fit Eq.~\eqref{eq:lorentzian} to the data near resonance and compare the fitted peak excitation $A$ with the numerically calculated on-resonance value.

We extract the Ramsey contrast from an ellipse fitted to the excitation fractions of the two ensembles using maximum-likelihood estimation, assuming a noise model limited by quantum projection noise.
We obtain the error bars by jackknife resampling, as described in detail in our previous work~\cite{Xin_multiplexedOLC}.

For the Ramsey coherence time measurements, we fit the data to an empirically motivated bi-exponential model,
\begin{equation}
C(t)=C_0\left[f e^{-t/T_f}+(1-f)e^{-t/T_s}\right].
\label{eq:double_exp}
\end{equation}
\noindent Here $f\in[0,1]$ represents the fraction of atoms undergoing the fast decay, $T_f$ and $T_s$ are the fast and slow decay times respectively, and $C_0$ is the contrast at zero dark time. We then define the $1/e$ coherence time as the time $t$ at which $C(t)= C_0/e$, determined numerically from the fitted Eq.~\ref{eq:double_exp}.

\subsection*{Empirical model for the coherence time}
As a guide to the eye and heuristic for the dependence of the Ramsey coherence time $\tau$ on the secondary lattice depth $U_{497}$ (Fig.~\ref{fig:4}B\&D), we fit the measured decay rates with the empirical function
\begin{equation}
\frac{1}{\tau} = \frac{\Gamma_{\mathrm{tun}}}{1+\left(U_{497}/U_{\mathrm{c}}\right)^{2}}
+ \Gamma_{0} + \gamma\,U_{497}.
\label{eq:tau_empirical}
\end{equation}
The first term describes the suppression of tunneling-induced decoherence with increasing secondary lattice depth, where $\Gamma_{\mathrm{tun}}$ is the decay rate in the absence of the secondary lattice and $U_{\mathrm{c}}$ sets the secondary lattice depth at which this contribution is halved.
$\Gamma_{0}$ captures depth-independent technical decoherence (e.g.\ magnetic field gradients and vector light shifts), and $\gamma$ captures an additional decay rate that grows linearly with the secondary lattice depth.
$\Gamma_{\mathrm{tun}}$ and $U_{\mathrm{c}}$ are fitted independently for each primary lattice depth ($10$, $20$, and 30\,\erecR), while $\Gamma_{0}$ and $\gamma$ are shared across all primary lattice depths.
This function is not derived from a microscopic model and is used only to capture the observed qualitative trends.

\subsection*{Optimal lattice depths in microgravity}

Here we estimate the optimal lattice depths for operating an optical lattice clock in microgravity, considering the fundamental limits placed on atomic coherence by tunneling, lattice induced Raman scattering, and the finite \tripletPzero lifetime. For a single 813\,nm lattice, increasing the lattice depth suppresses tunneling but increases the Raman scattering rate, resulting in an optimal operating depth set by the balance between these two effects. As shown in Fig.~\ref{fig:optimal_coherence}(A), the maximum combined $1/e$ coherence time is 36.6\,s at a lattice depth of approximately 100\,\erecR.

With the addition of the 497\,nm secondary lattice, Aubry--Andr\'{e} localization suppresses tunneling and the 813\,nm primary lattice can be operated at a substantially lower depth. For a secondary lattice depth of $U_{497}=0.133$\,\erecB, the optimal operating depth for the primary 813\,nm lattice shifts to $U_{813}=16$\,\erecR, where the calculated coherence time reaches 116.6\,s (Fig.~\ref{fig:optimal_coherence}(B)). This operating point is just above the Aubry--Andr\'{e} phase transition, where the maximum achievable coherence time is limited by Raman scattering from the two lattices and the \tripletPzero lifetime, rather than by tunneling.

\subsection*{Systematic offsets from the bichromatic superlattice clock}
In this section, we estimate the projected systematic shifts introduced by the addition of the secondary lattice. We consider two additional contributions arising from hyperpolarizability and frequency line pulling. We compare the optimal operating points of the single 813\,nm lattice and the bichromatic superlattice identified in the previous section. The lower primary lattice depth in the bichromatic configuration reduces the 813\,nm hyperpolarizability shift by approximately $40\times$ compared with the single 813\,nm lattice operated at 100\,\erecR.
\begin{figure}[ht!]
    \centering
    \includegraphics[width=1.0\linewidth]{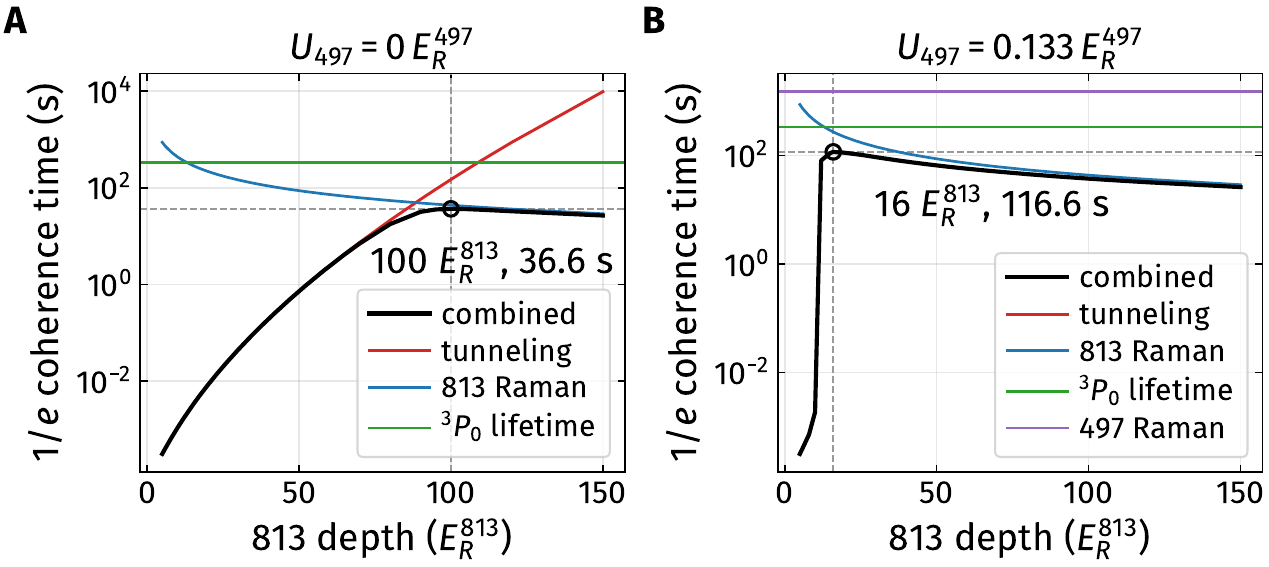}
    \caption{\textbf{Calculated coherence time with and without the secondary lattice.} Coherence time as a function of the 813\,nm lattice depth, showing the combined limit and the individual contributions from tunneling, Raman scattering, and the \tripletPzero lifetime. Circles mark the optimal operating points. (\textbf{A}) For a single 813\,nm lattice, the optimal coherence time of 36.6\,s is reached at $100\,\erecR$. (\textbf{B}) With a secondary 497\,nm lattice of $0.13\,\erecB$, the optimal coherence time is 116.6\,s at $16\,\erecR$.}
    \label{fig:optimal_coherence}
\end{figure}

However, the 497\,nm secondary lattice adds its own contribution to the hyperpolarizability shift in the superlattice.
To our knowledge, the hyperpolarizability at 497\,nm has not been measured or calculated, so we evaluate only the intensity factor $\expval{U_{497}^2}$, to which the shift is proportional for a given hyperpolarizability coefficient.
We calculate $\expval{U_{497}^2}$ assuming that the atoms uniformly occupy the lattice sites and remain in the vibrational ground state of the local potential, and compare it with $\expval{U_{813}^2}$.
At this operating point, $\expval{U_{497}^2}$ is approximately $2\times10^{-4}$ of $\expval{U_{813}^2}$. If the hyperpolarizability coefficient at 497\,nm is comparable in magnitude to that at 813\,nm, we therefore expect the additional hyperpolarizability shift from the secondary lattice to be at least three orders of magnitude smaller than that from the primary lattice.
Under the same assumptions, the standard deviation of the site-to-site variation of $\expval{U_{497}^2}$ is approximately 1.7 times smaller than that produced by the radial thermal distribution at a temperature of 1\,\uK, typical of optical lattice clock operation.
A complete treatment, including two-photon contributions, will require ab initio calculations or a measurement of the hyperpolarizability coefficients for the 497 nm light. 

\subsubsection*{Line pulling}
We estimate line pulling arising from the complex band structure of the superlattice using the same continuum supercell approach described above, extended to include all motional bands within the plane wave cutoff and finite temperature occupations.
We consider the projected optimal operating point with lattice depths of 16\,\erecR and 0.133\,\erecB, for which the motional Hamiltonian is
\begin{equation}
H_m =
\frac{p_z^2}{2m}
-16\,\mathrm{E_R^{813}} \,\sin^2(k_{813}z)
-0.133\,\mathrm{E_R^{497}}\sin^2(k_{497}z).
\end{equation}
Both lattices are assumed to operate at their respective magic wavelengths, such that the two clock states experience the same potential.

\begin{figure}[ht!]
    \centering
    \includegraphics[width=0.9\linewidth]{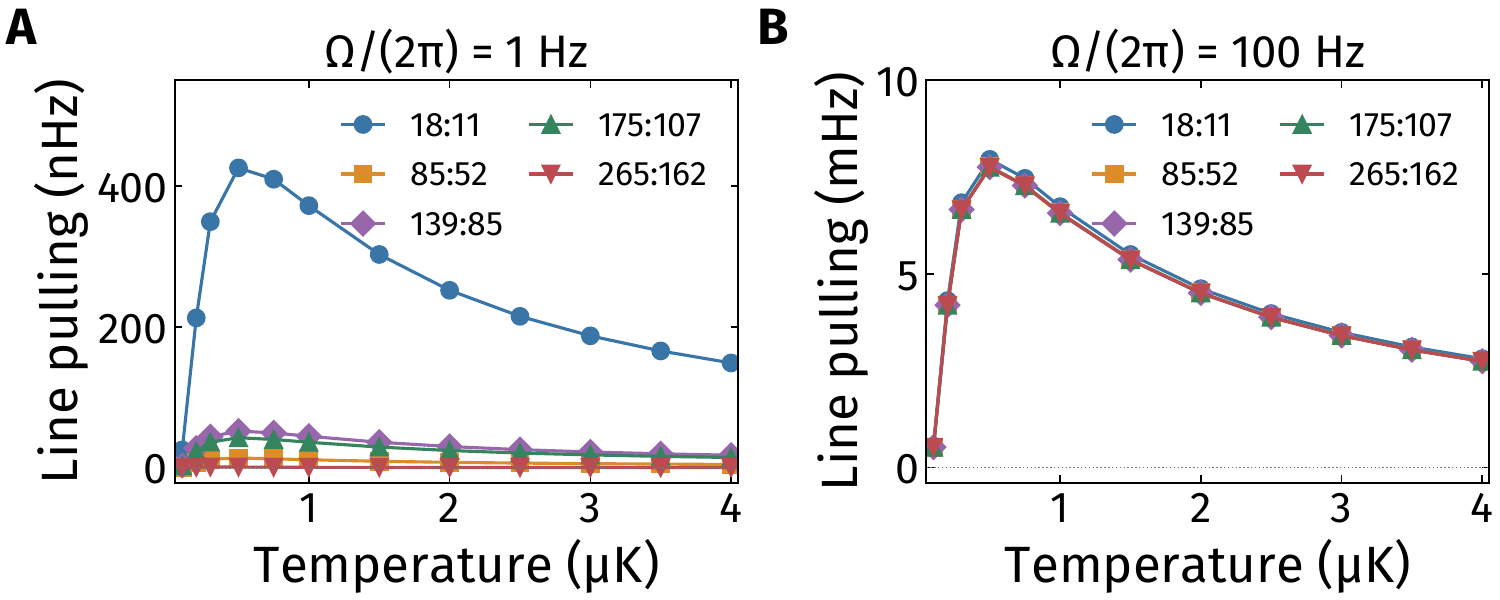}
    \caption{\textbf{Temperature dependence of motional line pulling in a bichromatic lattice.} The calculated shift of the carrier peak is shown for supercell wavelength ratios 18:11, 85:52, 139:85, 175:107, and 265:162. \textbf{A} and \textbf{B} correspond to Rabi frequencies $\Omega/(2\pi)=1$\,Hz and 100\,Hz, respectively.}
    \label{fig:S4}
\end{figure}
As in the simulation for Fig.~\ref{fig:3}, we approximate the wavelength ratio using rational supercells (18:11, 85:52, 139:85, 175:107, 265:162) and diagonalize the continuum Hamiltonian in a plane wave basis for each quasimomentum $q$.
For the line pulling calculation, we use a sequence of larger rational supercells to verify convergence with the supercell size.
The resulting eigenstates and eigenenergies are denoted by $\ket{m,q}$ and $\epsilon_m(q)$.
The initial motional states are populated according to a Fermi-Dirac distribution with unit filling in 813 lattice.

We calculate the excitation probability after a coherent $\pi$--pulse from the initial state.
The line pulling shift is defined as the displacement of the carrier maximum from zero detuning.
Numerically, we determine this shift using a local fourth order interpolation of the even and odd components of the calculated spectrum near resonance and locate the maximum closest to zero detuning.

Figure~\ref{fig:S4} shows the temperature and probe strength dependence of the calculated line pulling.
For $\Omega/(2\pi)=100$\,Hz, the results approach convergence as the supercell size increases.
For the largest supercell, the shift increases from 0.52\,mHz at 0.1\,\uK\ to 7.75\,mHz at 0.5\,\uK, and then decreases to 2.74\,mHz at 4\,\uK.
For $\Omega/(2\pi)=1$\,Hz, the results remain more sensitive to the supercell size, as expected from the longer interrogation time and the corresponding spreading of the atomic wavefunction.
Nevertheless, all calculated shifts are below 0.43\,\uHz, more than six orders of magnitude smaller than the Rabi frequency.

We can also understand the suppression of line pulling in the ideal limit of uniform site occupation within a single motional band, neglecting probe coupling to higher bands.
Although the bichromatic lattice does not possess the local spatial symmetry of a vertical lattice, each transition between motional states $a$ and $b$ has a corresponding reverse transition from $b$ to $a$.
The two transitions have opposite detunings and equal coupling strengths.
When the initial states are equally populated, they therefore contribute equal spectral weight at opposite detunings.
The ensemble spectrum is consequently symmetric about zero detuning, and no motional line pulling of the carrier occurs in this limit.

\end{document}